\documentclass[5pt]{article}

\usepackage[letterpaper]{geometry}
\usepackage{spconf,amsmath,epsfig,amssymb}
\newcommand{\norm}[1]{\left\lVert#1\right\rVert}

\usepackage{algorithm,algorithmic}
\floatname{algorithm}{Workflow}

\usepackage{flexisym}
\usepackage{enumitem}
\usepackage{siunitx}

\usepackage{graphicx}
\usepackage{subcaption}

\usepackage{color}
\usepackage{multirow}
\graphicspath{../figures/}

\usepackage{etoolbox}
\makeatletter
\patchcmd{\maketitle}{\@copyrightspace}{}{}{}
\makeatother


\usepackage{tabularx}

\newcounter{Workflow}

\usepackage{fancyhdr}
\begin{document}\sloppy

\def\x{{\mathbf x}}
\def\L{{\cal L}}

\title{From Thumbnails to Summaries - A single Deep Neural Network to Rule Them All}
%
%
%
\twoauthors
  {Hongxiang Gu\sthanks{The first author performed the work
	while at Adobe Research}}
	{University of California, Los Angeles\\
	Department of Computer Science\\
	Los Angeles\\
	hxgu@cs.ucla.edu}
  {Viswanathan Swaminathan}
	{Adobe Research \\
	BigData Experience Lab\\
	San Jose\\
        vishy@adobe.com}

\maketitle

\begin{abstract}
Video summaries come in many forms, from traditional single-image thumbnails, animated thumbnails, storyboards, to trailer-like video summaries. Content creators use the summaries to display the most attractive portion of their videos; the users use them to quickly evaluate if a video is worth watching. All forms of summaries are essential to video viewers, content creators, and advertisers. Often video content management systems have to generate multiple versions of summaries that vary in duration and presentational forms. We present a framework ReconstSum that utilizes LSTM-based autoencoder architecture to extract and select a sparse subset of video frames or keyshots that optimally represent the input video in an unsupervised manner. The encoder selects a subset from the input video while the decoder seeks to reconstruct the video from the selection. The goal is to minimize the difference between the original input video and the reconstructed video. Our method is easily extendable to generate a variety of applications including static video thumbnails, animated thumbnails, storyboards and "trailer-like" highlights. We specifically study and evaluate two most popular use cases: thumbnail generation and storyboard generation. We demonstrate that our methods generate better results than the state-of-the-art techniques in both use cases.
\end{abstract}
\section{Introduction}
\label{sec:intro}
Video summaries are widely used in many video related applications. Good video summaries serve the purpose of building anticipation while accurately representing the main content in the video. In order to maximize the probability of viewer clicks, two requirements are usually imposed during the process of searching for a good thumbnail generations: representativeness and aesthetics. A video summary should be representative of the original video to accurately convey the main theme to potential viewers. A good summary should also be clear, aesthetically pleasing and appealing to avoid confusion and attract view clicks. 

The advancement of human-computer interaction technology has enabled many variants of video summarizations. The most widely deployed ones are:
\begin{itemize}[topsep=0pt,itemsep=-1ex,partopsep=1ex,parsep=1ex]
\item Thumbnails. Video thumbnails are usually the first thing viewers see when browsing on a video website like YouTube. It is usually a single static image selected from the original video. If viewers are interested, they can click on the thumbnail to see the full video. Thumbnails allow viewers to have control over exactly what they want to see. 
\item Animated thumbnails. Animated thumbnails emerge on websites like Youtube as an improvement of single image thumbnails with a continuous short video clip (usually 2-3 seconds long). Animated thumbnails provide much more abundant information about the video while making "click baits" more obvious.
\item Storyboards. To strengthen the ability to represent the video content, some video platform allows viewers to quickly scan through the entire video by presenting multiple keyframes selected from the original video and present them as storyboards.
\item Trailer-like summaries. Some video websites concatenate multiple key shots to create a "trailer-like" short video which provides much richer content comparing to storyboards.
\end{itemize}

\begin{figure}[h!]
    \centering
    \includegraphics[width=2.5in]{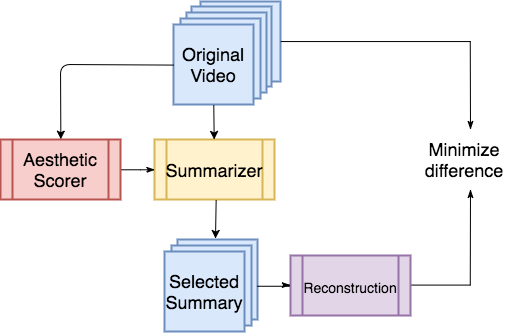}
    \caption{Overview of proposed video summarization through video reconstruction. The summarizer takes consideration of both aesthetic factors and temporal dependency between frames. }
    \label{fig:overview}
\end{figure}

Various forms of videos summaries require different mechanisms to generate. Many thumbnail generation algorithms perform excellently in finding a single image to represent the input video but fail to capture temporal information in the video, thus not suitable for storyboard or "trailer-like" summary generation. On the other hand, many mechanisms dedicated to generating storyboards or "trailers" are incapable of finding one single representative image. In this work, we present a single deep learning framework that can be utilized to generating summaries in different forms. 

Our main idea is inspired by the intuition that if a sparse subset of the video frames can be reconstructed to a new video that has a minimum difference from the original video, then the selected subset is the most representative selection. The selection is done by assigning normalized importance scores to each frame in the original video while the reconstruction is done through an LSTM-based autoencoder network. The original video is being weighted by the frame by frame importance score through multiplication merge. Apart from relevancy considerations, a pre-trained CNN-based aesthetic scorer trained on AVA dataset ensures that the selected frames used for summary generation are not only representative but also clear and aesthetically pleasing. An overview of our core idea is shown in figure \ref{fig:overview}.

We demonstrate how our work can be utilized to generate high-quality thumbnails, animated thumbnails, storyboards and "trailer-like" summaries by simply changing the selector regularizers.  More specifically, we show that our work provides better results when comparing to the state-of-the-art video summarization mechanism in two popular use cases: thumbnails and storyboards. Animated thumbnails and "trailer-like" summaries are not evaluated due to lack of comparison in existing literature, but we present an efficient workflow in generating these two summaries and three demos in the supplementary materials.

\section{Related Work}
\subsection{Automated Thumbnail Selection}
Video thumbnails are most compact-size versions of videos to capture the essence of a video and give out first impressions to potential viewers. They are usually presented in the form of a single image. Traditionally, many thumbnails are selected by humans which is expensive and unsatisfactory.  Much research has been conducted in automating the process of thumbnail generation in the past. In order to improve relevancy, Gao et al. proposed to utilize semantic information to select semantically representative frames as video thumbnails \cite{gao2009thematic}. Liu et al. developed a multi-task deep visual-semantic embedding model to automatically select query-dependent thumbnails based on both semantic and visual features\cite{liu2015multi}. Both methods heavily depend on using semantic information to guarantee the representativeness of the selection. However, in a real-world scenario, there is no assurance of the quality of semantic information. False or meaningless titles, descriptions or audio tracks could jeopardize the quality of the selected thumbnails. In this paper, we assume that no semantic information is available thus we use only visual features of video frames and the temporal relationship among them. 

In addition to improving relevancy, Song et al. presented work on selecting not only relevant but also aesthetically pleasing thumbnails by utilizing an aesthetic scoring mechanism jointly with K-nearest neighbor algorithm \cite{song2016click}. We adopt a similar idea by using a pre-trained aesthetic scorer to eliminate unclear or blurry thumbnails. Our aesthetic scoring model is trained on AVA dataset and is directly applied to the attention module to select clear and aesthetically pleasing thumbnails.

\subsection{Video Summary Generation}
Video summarization has been studied in both academia and industry for many years due to its importance in video understanding, video management, and digital marketing.  In 2007, Truong et al. surveyed eight different video summarization mechanisms \cite{truong2007video} including: sufficient content change detection, equal temporal variance, maximum frame coverage, clustering etc. 
As we step into the era of machine learning, numerous efforts have been made in using machine learning techniques to summarize videos. Zhang et al. proposed to use Long Short-Term Memory (LSTM) to model the variable-range temporal dependency among video frames in a supervised manner \cite{zhang2016video}. However, due to the limited number of annotated videos, it is questionable how supervised learning would perform on much larger and more diverse datasets. To address this problem, Mahasseni et al. proposed an unsupervised video summarization method that utilizes adversarial LSTM networks \cite{mahasseni2017unsupervised}. The main idea is to minimize the discrimination between the deep features of the original video and that of a selected subset of frames. Even though the solution achieves outstanding results in four popular benchmark datasets, adversarial training is computationally hungry and unstable. Also, the solution does not take selection quality into consideration, often selects blurry or transitional frames which cannot be directly used in real-world applications.

\section{Our Contribution}
Comparing to previous works, we highlight the following characteristics of our work:
\begin{enumerate}
\item We build a system that can be used to generates video summary of many formats including but not limited to video thumbnails, animated thumbnails, storyboards and "trailer-like" video summaries with one unsupervised deep learning model.
\item Our summarization considers both aesthetics and relevancy in selecting keyframes to make resulting summary accurate and appealing. Our results outperforms state-of-the-art unsupervised methods in frame level and keyshot level evaluation.
\end{enumerate}

\section{Proposed Framework}
\begin{figure}[h!]
    \centering
    \includegraphics[width=2.5in]{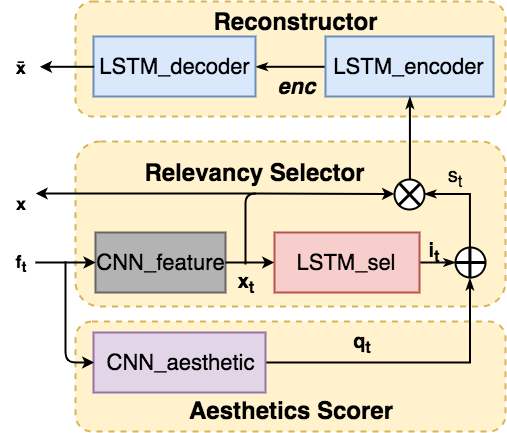}
    \setlength{\belowcaptionskip}{-12pt}
    \caption{Main components of our framework. The aesthetic scorer assigns aesthetic scores $q_t$ $\in$ [0,1] to each individual frame image $f_{t}$ from the original video. The relevance selector selects a subset of frames from the input sequence $x$ and assigns an important score $i_t$ $\in$ [0,1] to each frame. The LSTM-based encoder encodes the selected frames into a fixed length feature vector $enc$ and then reconstructed the video sequence $\bar{x}$ in the decoder.}
    \label{fig:framework}
\end{figure}

Our proposed framework ReconstSum consists of three major components: the aesthetic scorer, the relevance selector, and the reconstructor, as illustrated in Figure \ref{fig:framework}.

\subsection{Aesthetic Scorer}
Our proposed framework utilized an aesthetic scorer to select clear and aesthetically pleasing images as thumbnails or other formats of video summary. The aesthetic scorer is created by fine-tuning the fully-connected layers of a pre-trained InceptionV3 network \cite{szegedy2016rethinking} on a large-scale visual aesthetic (AVA) dataset \cite{murray2012ava} using the techniques proposed by Jin et al. \cite{jin2016deep}. The aesthetic scorer takes a frame from the original video as input and generates an aesthetic score $q_t$ $\in$ [0,1], where $q_t = 0$ indicates an image of a poor aesthetic score and $q_t = 1$ indicates an image of the highest quality. In the extreme case of discretized aesthetic scores, an image is either qualified or not qualified for selection ($q_t$ $\in$ \{0,1\}).

\subsection{Relevance Selector}
The selector takes a sequence of deep features of every frame of the original video $x = \{x_t: t = 1,...,N\}$ as input. The deep features are extracted using a pre-trained InceptionV3 model (output of the global pooling layer of 2,048 dimensions). The selector by nature is a bidirectional LSTM. The selector generates a one dimensional vector of normalized importance score $i = \{i_t \in [0,1]: t = 1,...,N\}$ for each frame. Similar to aesthetic scores, an discretized version can also be generated. For the consideration of both relevancy and aesthetics, we creates a new metric call selection score $s = \{s_t \in [0,1]:  s = 1,...,N\}$ which is the linear combination of both importance score and aesthetic score $s_t = \alpha i_{t} + \beta q_{t}$. The feature vector for each single frame is weighted by the selection score. We varied the value of alphas (weight for aesthetic scores) and betas (weight for relevancy scores) in our experiment under five settings: alpha, beta = {(0,1), (0.25,0.75), (0.5,0.5), (0.75,0.25), (1,0)}. We observe that alpha = 0,25 and beta = 0.75 provides the best result, and we used this setting in our evaluation sections.

Note that this design enables the selector to be compatible with multiple metrics. In case that high-quality semantic information is available, a semantic model can also be introduced and contribute to the selection score using linear combinations.

\subsection{Reconstructor}
The reconstructor is an LSTM autoencoder that consists of a bidirectional LSTM encoder and a bidirectional LSTM decoder. Srivastava et al. showed that LSTM-based autoencoder is a powerful model for learning and representing video representations \cite{srivastava2015unsupervised}.  The encoder takes the whole sequence of deep features of each video frames as input. The state of the encoder after the last feature vector has been taken is the full representation of the input sequence. Bear in mind that once we have the video representation of the selected subset of the original video, the decoder tends to reconstruct the video from the compressed representation. In \cite{srivastava2015unsupervised}, the reconstruction process might learn a direct identical mapping from the input to the output. We do not have this concern since our input is a weighted sequence of the original video whereas the target of the reconstruction is the sequence of the original videos. Thus a full reconstruction is not likely if the selection is sparse.

\subsection{Training The Network}
In our implementation, our training is defined by two loss functions: 
\begin{enumerate}
\item $\mathcal{L}_{reconst}$ is the reconstruction loss function for the reconstructor. In previous works that utilize LSTM-based autoencoders to predict video frames, authors conclude that the choosing the right loss function is extremely important and the squared loss function suffers from some drawbacks \cite{ranzato2014video}. They claim that squared loss function is not sensitive to minor distortions in the input sequence, thus does not provide optimal results in the training process. In  \cite{mahasseni2017unsupervised}, the authors used a discriminator as a replacement of squared loss functions. However, using such a discriminator can be expensive and difficult. Adversarial training is known for its difficulty in training, during our implementation of the framework proposed by \cite{mahasseni2017unsupervised}, we find that training in practice is oscillatory, often results in instability in resulting quality. Also using adversarial training is much more time and energy consuming. We conclude that in real applications, using squared loss function is much more efficient and sufficient enough to generate competitive results comparing to using other loss functions. We use regular squared loss functions to minimize the difference between $x$ and $\bar{x}$ in Figure \ref{fig:framework}

\item $\mathcal{L}_{sparsity}$ is the sparsity loss function for the selector. When the selector is not regularized by a sparsity function, the relevance selector would simply select every single frame to minimize the reconstruction loss. We have three variants of $\mathcal{L}_{sparsity}$. 
\end{enumerate}
The first regularizer is described by equation \ref{eq:sparsity1}. 
\begin{eqnarray}
\mathcal{L}_{sparsity} = \norm{\sum_{t=1}^N (i_t-\delta) } + \sum_{t=1}^N  entropy(i_t)
\label{eq:sparsity1}
\end{eqnarray}

The first term penalizes the action where the relevance selector selects many frames. N is the total number of frames of the original video, $i_t$ is relevance score of the t-th frame, $\mathbf{i}$ is the selection vector and $\delta$ is a parameter indicating proportion of maximum number frames can be selected to N. We discovered with the first term alone sometimes result in the relevance selector giving a uniform score of $\frac{\delta}{N}$ to all frames. Thus, we use the second entropy term to encourage strong opinions. The second term calculates the entropy of the entire selection vector. We want the selector to obtain a strong opinion on either to select or not to select a frame.

One drawback of this sparsity regularizer is that the result selection often contains similar looking frames. While this is not a problem in generating thumbnails or animated thumbnails, it does not provide sufficient diversity in more complicated applications like storyboards or "trailer-like" video summary. 

In order to address this problem, we use a repelling regularizer proposed by Zhao et al. in their EBGAN autoencoder model \cite{zhao2016energy}. The regularizer is described in equation \ref{eq:sparsity2}. 

\begin{eqnarray}
\mathcal{L}_{sparsity}^{rep} = \frac{1}{N(N-1)} \sum_{t}\sum_{t' \neq t} (\frac{h^{\intercal}_{t}h_{t'}}{\norm{h_t}\norm{h_{t'}}})^2
\label{eq:sparsity2}
\end{eqnarray}

where $h_t$ is the hidden state of the LSTM$_{enc}$ at time t. The repelling regularizer penalizes selecting from data in clustered together and attempts to orthogonalize the pairwise sample representation in the selection. In other words, the repelling regularizer encourages diversity in selection. 

\section{Case Study}
We demonstrate the process of generating high-quality thumbnails and storyboard summaries using our framework in this section. We also show that ReconstSum outperforms state-of-the-art techniques regarding quality and latency on three popular datasets: Summe \cite{GygliECCV14}, OVP, and Youtube \cite{Avila}. We evaluate our work at two levels. 

At frame level, we adopt the classic top-K evaluation methods by calculating the possibility of the top-K generated summaries match the top-K human selections. To note that often our method and human judges select visually similar but not the same frame in the video. To make the evaluation more efficient and more convincing, we consider our selection and the human selection a match if the Structural Similarity Index (SSIM) score between them is greater than 0.7.

At keyshot level, we use the evaluation method proposed by Zhang et al. We use video segmentation methods to find two sets of keyshots that are selected by our framework (A) and by human judges (B). The accuracy of the summarization is calculated as the harmonic mean F-score.

\begin{eqnarray}
P = \frac{\textnormal{A} \cap \textnormal{B}}{\norm{\textnormal{A}}}, R = \frac{\textnormal{A} \cap \textnormal{B}}{\norm{\textnormal{B}}} \\
F = \frac{\textnormal{2P} \times \textnormal{R}}{\textnormal{R(P + R)}}
\label{eq:PR}
\end{eqnarray}

When evaluating thumbnail generation, we use only frame level evaluation as no keyshot is involved. For storyboard generation, we evaluate our method at both frame and keyshot level. We evaluate variants of our framework including: ReconstSum where only regularizer described by equation \ref{eq:sparsity1} is used, ReconstSum$_{rep}$ where both regularizers described by equation \ref{eq:sparsity1} and  \ref{eq:sparsity2} are used and ReconstSum$_{disc}$ where both regularizers are used and the selector generates discretized outputs.

\subsection{Thumbnail}
To generate $m$ candidate thumbnail using our framework, we simply set the parameter $\delta$ in the sparsity regularizer (equation \ref{eq:sparsity1}) to be $\frac{m}{N}$. We train the autoencoder until convergence and extract the selection from the selector.

As all three datasets do not directly contain thumbnail information, we use the top-3 most selected frames selected by all human judges as the top-3 candidate thumbnails for each video. 

\begin{table}[t]
\begin{center}
\caption{Comparison of different variations of our thumbnail selection with the state of the art for SumMe and TVSum datasets with videos from OVP and YouTube data using top-3 matching evaluation.} \label{tab:cap}
\begin{tabular}{|c|c|c|}
  \hline
  Method & OVP & Youtube
  \\
  \hline
  \cite{mahasseni2017unsupervised}\footnotemark &  7.80\% & 11.34\%   \\
  \cite{song2016click} &  11.72\% &  16.47\%\\
  ReconstSum &  9.06\% &  17.02\%\\
  ReconstSum$_{rep}$ &  11.84\% &  18.12\%\\
  ReconstSum$_{disc}$ &  12.18\% &  18.25\%\\
  \hline
\end{tabular}
\label{tb:top3}
\end{center}
\end{table}
\footnotetext{We repeat the implementation described by the authors as no original code was provided. Our implementation is verified by repeating some of the original experiments.}

Our results using top-3 evaluation are presented in Table \ref{tb:top3}. We observe that our implementation of \cite{mahasseni2017unsupervised} performs the worst and ReconstSum$_{disc}$ performs the best. When examining the selected frames, we observe that \cite{mahasseni2017unsupervised} often selects frames that contain transitional scenes or images of low aesthetic quality. Since human judges almost never select frames with low aesthetic qualities, frames selected by \cite{mahasseni2017unsupervised} fails to compete with other frameworks that consider aesthetics. One interesting observation is that ReconstSum$_{rep}$ outperforms ReconstSum as the repelling regularizer encourages diversity in candidate selection. Among all top 3 thumbnail candidates, 37\% of ReconstSum's selections contains at least two similar looking candidates (SSIM score higher than 0.7) where ReconstSum$_{rep}$ significantly reduces the number to only 4\%. ReconstSum$_{disc}$ performs better than the non-discretized version of the selector in thumbnail selection; we believe it is because a discretized aesthetic scorer further eliminates the candidacy of low-quality frames, thus making the model behaves more like human. 

\subsection{Story Boards}
In order to have a fair comparison between the storyboards generated by ReconstSum and other previous works, we adopt the keyshot evaluation method used in many recent works \cite{zhang2016video}\cite{mahasseni2017unsupervised}. 

\begin{table}[t]
\begin{center}
\caption{F-score comparison of storyboard generated by our proposed approach to state-of-the-art at keyshot level. The reported results from the state of the art are from published results. } 
\begin{tabular}{|c|c|c|c|}
  \hline
  Method & Summe & OVP & Youtube\\
  \hline
  \cite{furini2010stimo} &  - &  63.4 & - \\
  \cite{Avila} &  33.7 &  70.3 & 59.9 \\
  \cite{mahasseni2017unsupervised} &  39.1 & 72.8  & 60.1  \\
  \cite{mahasseni2017unsupervised}$^1$ &  37.9 & 71.9  & 60.3  \\
  ReconstSum$_{rep}$ &  39.8 &  71.7 & 61.5 \\
  \hline
\end{tabular}
\label{tb:keyshot}
\end{center}
\end{table}

Table \ref{tb:keyshot} summarizes the accuracy of storyboards generated by our approach. Our ReconstSum$_{rep}$ outperforms all state-of-the-art techniques on all three dataset except \cite{mahasseni2017unsupervised} on OVP dataset. 

When using storyboards in real applications, however, keyshot based evaluation is not sufficient enough as an indicator of the storyboard quality. Instead, we also care about which specific images are presented to the viewers. Again, we use top-K evaluation method. This time we set 
\begin{eqnarray}
K = min(\textnormal{len}(\cup_1^n u_i), \textnormal{len}(sb))
\label{eq:K}
\end{eqnarray}
where $u_i$ is the storyboard selected by each human judge, n being the total number of human judges, and $sb$ being our generated storyboard. Table \ref{tb:topk} displays our evaluation results on storyboard generation at frame level. ReconstSum$_{rep}$ outperforms state-of-the-art work and has the highest top-K accuracy among all variants of our proposed framework. Even though ReconstSum$_{rep}$ and \cite{mahasseni2017unsupervised} both performs competitively at keyshot level, ReconstSum$_{rep}$ completely dominates \cite{mahasseni2017unsupervised} by 4.15\%, 49.89\% and 54.18\% at frame level for all three datasets. Noted that all methods generate high top-K accuracy on Summe dataset since benchmark videos have very little scene changes, resulting in all selected images being highly similar (high SSIM scores).

\begin{table}[t]
\begin{center}
\caption{Top-K accuracy comparison of storyboard generated by our proposed approach to state-of-the-art at frame level. } 
\begin{tabular}{|c|c|c|c|}
  \hline
  Method & Summe & OVP & Youtube\\
  \hline
  \cite{Avila} & 85.25\%  &  22.19\% & 24.6\% \\
  \cite{mahasseni2017unsupervised}$^1$ & 85.35\% & 19.24\%  & 19.47\%  \\
  ReconstSum &  84.40\% &  24.22\% & 25.12\% \\
  ReconstSum$_{disc}$ &  87.25\% &  26.96\% & 27.98\% \\
  ReconstSum$_{rep}$ &  88.89\% &  28.84\% & 30.02\% \\
  \hline
\end{tabular}
\label{tb:topk}
\end{center}
\end{table}

\subsection{Animated Thumbnails and Trailer-like Summary}
Both animated thumbnails and trailer-like summary are supported by more and more video websites like Youtube. When combined with proper video segmentation techniques like Kernel Temporal Segmentation (KTS)  \cite{potapov2014category}, our framework can be adapted to the production of both summary formats using Workflow \ref{wf}:
\begin{algorithm}[!ht]
\textit{Inputs.} Source video $V$ containing $n$ frames.

\textit{Output} Video Summary $Sum$.

\begin{enumerate} [topsep=0pt,itemsep=-1ex,partopsep=1ex,parsep=1ex]
	\item \textbf{Video segmentation}
    		Slice video into N segments, $N \leq n$; return Segmentation $Seg$
	\item \textbf{Selection}
		Use \textit{ReconstSum} to select $m$ frames (m=1 if generating animated thumbnails); return the selection vector $Sel$.
	\item \textbf{Score assignment}
		Assign scores to each segment $Seg_i$. For $i \leq N$, $j \leq n$:
		\begin{equation}
		\text{score}(Seg_i) = \sum_{j}^{\textnormal{j+len}(Seg_i)} Sel_j, j += len(Seg_i).
		\end{equation}
	\item \textbf{Summary generation}
		$Sum$ = knapsack($Seg$, $L$) to maximize value($Sum$),  $Sum$ $\subset$ $Seg$.
\end{enumerate}
\caption{}\label{wf}
\end{algorithm}



\section{Conclusion and Future Work}
Our work explores a single LSTM-based autoencoder structure that is capable of selecting most representative and aesthetically pleasing summary in an unsupervised manner. The main objective is to use an LSTM-based selector and an aesthetic scorer to select a sparse subset of frames so that the reconstructed from the selection has a minimum difference with the original video. We have also shown quantitively that our model outperforms state-of-the-art unsupervised video summarization techniques by 3.92\%-10.8\% in thumbnail selection and by at least 4.15\% in storyboard generation. Lastly, we have also designed a workflow for generating animated thumbnails and trailer-like summaries utilizing our framework. Unfortunately, lack of annotated data on animated thumbnails and trailer-like summaries has limited our ability to further evaluate the quality of our proposed workflow. We intend to create a benchmark dataset and new evaluation methods to quantitively measure the quality of our proposal in the future.

\section{Citations and References}

\bibliographystyle{IEEEbib}
\bibliography{icme2018hxgu}

\end{document}